\documentclass[prd,twocolumn,amsmath,amssymb,nofootinbib,floatfix,superscriptaddress]{revtex4}

\usepackage{slashed} 
\usepackage{physics, xcolor, bbold}
\usepackage{graphicx,bm}
\usepackage[normalem]{ulem} 

\makeatletter
\def\graphicscale{\twocolumn@sw{0.3}{0.4}}
\def\graphicthreescale{\twocolumn@sw{0.3}{0.4}}

\begin{document}

\title{Chiral critical behavior of 3D lattice fermionic models with
  quartic interactions 
}

\author{Claudio Bonati} 
\affiliation{Dipartimento di Fisica dell'Universit\`a di Pisa 
  and INFN Largo Pontecorvo 3, I-56127 Pisa, Italy}

\author{Alessio Franchi} 
\affiliation{Dipartimento di Fisica dell'Universit\`a di Pisa 
       and INFN Largo Pontecorvo 3, I-56127 Pisa, Italy}

\author{Andrea Pelissetto}
\affiliation{Dipartimento di Fisica dell'Universit\`a di Roma Sapienza
        and INFN Sezione di Roma I, I-00185 Roma, Italy}

\author{Ettore Vicari} 
\affiliation{Dipartimento di Fisica dell'Universit\`a di Pisa
       and INFN Largo Pontecorvo 3, I-56127 Pisa, Italy}

\date{\today}

\begin{abstract}

We study the critical behavior of the three-dimensional (3D)
Gross-Neveu (GN) model with $N_f$ Dirac fermionic flavors and quartic
interactions, at the chiral ${\mathbb Z}_2$ transition in the massless
${\mathbb Z}_2$-symmetric limit.  For this purpose, we consider a
lattice GN model with staggered Kogut-Susskind fermions and a scalar
field coupled to the scalar bilinear fermionic operator, which
effectively realizes the attractive four-fermion interaction.  We
perform Monte Carlo (MC) simulations for $N_f=4,8,12,16$. By means of
finite-size scaling (FSS) analyses of the numerical data, we obtain
estimates of the critical exponents that are compared with the
large-$N_f$ predictions obtained using the continuum GN field
theory. We observe a substantial agreement.  This confirms that
lattice GN models with staggered fermions provide a nonpertubative
realization of the GN quantum field theory, even though the lattice
interactions explicitly break the flavor ${\rm U}(N_f)\otimes {\rm
  U}(N_f)$ symmetry of the GN field theory, which is only recovered in
the critical limit.

\end{abstract}

\maketitle


\section{Introduction}
\label{intro}

Three-dimensional (3D) quantum field theories (QFTs) of interacting
fermions emerge in different contexts; for instance, in
condensed-matter physics, they are used to describe the low-energy
excitations in graphene, see, e.g.,
Refs.~\cite{NGPNG-09,VZS-00,KLOSFK-08,HJV-09,AH-13,PHA-15,OYS-16,
  XG-20,OSSY-20}. Among them, we should mention quantum
electrodynamics with charged fermions, the Gross-Neveu and the
Gross-Neveu-Yukawa models, in which the dynamics of Dirac fermions
arises from four-fermion interaction terms or through the coupling
with a scalar field~\cite{ZJ-book}.

In this paper, we focus on the Gross-Neveu (GN) QFT defined by the
Euclidean Lagrangian density
\begin{eqnarray}
{\cal L} = -\sum_{f=1}^{N_f} \bar{\Psi}_f(\slashed{\partial}
    + m)\Psi_f - \frac{g^2}{2N_f}\, \Big(\sum_{f=1}^{N_f} \bar{\Psi}_f
  \Psi_f\Big)^2\,,
  \label{lagrangianGN}
\end{eqnarray}
where $\Psi_f({\bm x})$ with $f=1,\ldots, N_f$ is a fermionic field.
Each {\em flavor} component $\Psi_f$ is a four-dimensional spinor, so
that the total number $N$ of fermionic components is given by $N = 4
N_f$, and the matrices $\gamma_\mu$ are the usual Euclidean 4$\times$4
matrices used in 4 dimensions \cite{BB-87}.  This choice allows us to
define chiral symmetry transformations~\cite{P-84}. It is also somehow
necessary if we wish to compare our findings with $\epsilon$-expansion
($\epsilon = 4-d$) results obtained in the four-dimensional model with
standard Dirac spinors \cite{ABKW-86}.  Note that in three
  dimensions it is also possibile to define GN QFTs with two-component
  spinors: in this case no chirality is present, but an analogous role
  is played by the reflection with respect to one of the
  axes~\cite{ZJ-book,MZ-03}. The Lagrangian in
Eq.~\eqref{lagrangianGN}, with attractive four-fermion interactions,
can be equivalently written as
\begin{eqnarray}
 {\cal L} = -\sum_{f=1}^{N_f} \bar{\Psi}_f
    (\slashed{\partial} + m + g \,\Phi)\Psi_f
   + \frac{N_f}{2} \Phi^2\,,
     \label{lagrangianGN2}
\end{eqnarray}
where $\Phi({\bm x})$ is an auxiliary real scalar field associated
with the bilinear fermionic operator $\sum_{f} \bar{\Psi}_f
\Psi_f$.  Indeed, by integrating out the scalar field $\Phi$, one
recovers Lagrangian \eqref{lagrangianGN}.

The global flavor symmetry of the GN QFT is ${\rm U}(N_f)\otimes {\rm
  U}(N_f)$~\cite{BB-87,ABKW-86}. If fermions are massless, the 3D
Lagrangian is also invariant under two additional ${\mathbb Z}_2$
chiral transformations~\cite{BB-87,ABKW-86}
\begin{eqnarray}
&& \Psi_f \to \gamma_5\Psi_f\,,\quad \bar\Psi_f \to - \bar\Psi_f
\gamma_5\,,\quad \Phi \to - \Phi\,,  \nonumber \\
&& \Psi_f \to \gamma_4\Psi_f\,,\quad \bar\Psi_f \to - \bar\Psi_f
\gamma_4\,,\quad \Phi \to - \Phi\,.  
\label{z2symm}
\end{eqnarray}
The presence of two chiral symmetries is related to the fact that only
the $\gamma_\mu$ matrices with $\mu=1,2,3$ appear in the Lagrangian,
so that $\gamma_4$ and $\gamma_5$ play essentially the same
role~\cite{ABKW-86}.  In the massless GN models with attractive
interactions and at least for a sufficiently large number of
flavors~\cite{MZ-03}, there is a phase transition where the chiral
${\mathbb Z}_2$ symmetries are spontaneously broken. The transition
separates a disordered phase, in which the scalar-field correlations
are short-ranged and fermions are massless, from an ordered phase, in
which the scalar field orders (in field theory terms, it has a
nonvanishing expectation value), providing an effective mass for the
fermion fields.  As discussed in Refs.~\cite{ZJ-book,MZ-03}, a similar
behavior is expected in the Gross-Neveu-Yukawa (GNY) QFT, which is an
extension of the GN model obtained by adding a kinetic and a
quartic-interaction term for the real scalar field.

The massless GN QFT with attractive interactions should provide the
effective description of the critical behavior of systems with the
same global symmetry, symmetry breaking pattern and field content.  In
the case of the GN model, the symmetry that is broken is always the
chiral ${\mathbb Z}_2\otimes {\mathbb Z}_2$ symmetry, for any $N_f$.
Therefore, the relevant symmetry and symmetry-breaking pattern at the
transition are always the same.  However, the resulting critical
behavior depends on $N_f$, because of the different fermion content of
the model. Indeed, the chiral transition occurs in the presence of
$N_f$ massless fermions, which generate long-range interactions for
the scalar field, which obviously depend on $N_f$.

The renormalization-group (RG) flow in the GN and GNY QFTs has been
investigated using different methods. Critical exponents have been
computed in the $d$-dimensional theory in the large-$N_f$
limit. Results to order $1/N_f^2$ are reported in
Refs.~\cite{MZ-03,HKK-92,VS-93,VDKS-94,Gracey-93,Gracey-94,Gracey-94b,Gracey-17}.
They provide quantitative information that can be compared with
results obtained in statistical models that, supposedly, have
transitions associated with these QFTs.  The RG flow has also been
studied in perturbation theory. Perturbative calculations have been
performed around four dimensions in the GNY model~\cite{ZMMHS-17} to
four loops, providing the $\epsilon$ expansions of the critical
exponents up to $O(\epsilon^4)$. {\em A priori}, it is not clear if
these results directly apply to the three-dimensional model. Indeed,
in the four-dimensional model the Lagrangian is invariant under a
single chiral symmetry---the chiral symmetry group is ${\mathbb Z}_2$
---while in three dimensions the chiral symmetry group is larger,
being ${\mathbb Z}_2\otimes {\mathbb Z}_2$. However, the large-$N_f$
expressions of the critical exponents do not show nonanalyticites as
$d\to 3$, indicating that this dimension-dependent symmetry
enlargement should have no impact on the $d$-dependent analyticity
properties of the universal features of the model.  We should also
remark that $\epsilon$ expansions for the GNY model are not Borel
summable, at variance with what happens for the corresponding
expansions in $\Phi^4$ scalar theories, see, e.g.,
Refs.~\cite{ZJ-book,GZ-98,PV-02}. Therefore, we do not expect them to
provide accurate 3D estimates. A thorough analysis of the perturbative
series is reported in Ref.~\cite{IMS-18}. Exponents have also been
computed using the functional renormalization group \cite{RVW-01} and
the conformal-bootstrap approach~\cite{IKPPS-18,EIKLPS-22}.

We mention that numerical results for relatively small values of
$N_f$, i.e., $N_f\le 4$, have been reported in
Refs.~\cite{CS-07,CL-12,CL-13,HC-17,HXSAML-18,LWSM-20,OYS-21,MNMV-21},
using a variety of different formulations. These results have been
compared with the estimates obtained in the field-theory approaches,
(see, e.g., the results reported in Table 3 of Ref.~\cite{IKPPS-18},
where $N=4N_f$).  In some cases, large discrepancies are observed
among the results obtained [for instance, for $N_f=1$, the estimates
  of $1/\nu$ vary between 0.76 and 1.30(5)]. In particular, the
conformal-bootstrap results of Ref.~\cite{IKPPS-18}, which have been
recently confirmed in Ref.~\cite{EIKLPS-22}, provide estimates that
differ significantly from those obtained using numerical methods.
This uncertain situation calls for further studies, to understand
  the reasons of such discrepancies, whether and how the QFT scenarios
  get realized in the phase diagram of corresponding statistical
  lattice systems.

In this paper we investigate the critical behavior of 3D statistical
fermionic models defined on cubic lattices, to shed light on the way,
or whether, they realize the continuum GN QFT at the chiral
transition.  The definition of fermionic lattice models is affected by
the well-known fermion doubling problem~\cite{MM-book,CER-83}, which
makes it impossible to implement the quartic fermion interaction, or,
equivalently, the interaction between fermionic and scalar fields,
preserving the flavor symmetry ${\rm U}(N_f)\otimes {\rm U}(N_f)$.  A
partial solution is provided by Kogut-Susskind (KS)
formulations~\cite{CER-83,HKK-92}. In this case, two doublers are
present in the model, so that $N_f/2$ KS fermion variables per site
are needed to describe a system with $N_f$ flavors.  In these models
the Hamiltonian is only exactly invariant under $U(N_f/2)$ global
transformations and, in the massless limit, under a single ${\mathbb
  Z}_2$ chiral symmetry. If the scalar field variables are located on
the dual lattice sites~\cite{CER-83,HKK-92}, the terms that break the
symmetry between the doublers, and therefore the full flavor symmetry
of the continuum model, are $O(a)$ ($a$ is the lattice spacing) in the
formal {\em classical} continuum $a\to 0$ limit.  Therefore, the
symmetry ${\rm U}(N_f)\otimes {\rm U}(N_f)$ of the continuum GN and
GNY models is recovered at the leading {\em classical} tree order.  In
the RG context, this result is taken as an indication that these
breaking terms are irrelevant perturbations of the GNY QFT fixed
point, so that the lattice systems recover the continuum ${\rm
  U}(N_f)\otimes {\rm U}(N_f)$ symmetry at the chiral transition
point.  Thus, their asymptotic critical behavior belongs to the same
universality class as that of the continuum QFT of GN and GNY models.

Here we return to this issue, verifying whether the conjectured
irrelevance of the $O(a)$ flavor symmetry violations holds at the
chiral transition of 3D lattice GN-like models.  Indeed, although the
fact that the flavor-symmetry violating terms are $O(a)$ is a
necessary condition for the recovery of the full flavor symmetry, it
may not be sufficient at a nonperturbative level. Therefore, an
accurate check at a nonperturbative level is called for, carefully
studying the critical behavior at the chiral transition. For this
purpose, we focus on the large-$N_f$ regime and compare Monte Carlo
(MC) results with the available nonperturbative large-$N_f$ expansions
of the critical exponents computed using the GN QFT.  Since the
critical behavior in the $N_f\to\infty$ limit of lattice models
matches that of the continuum GN models~\cite{HKK-92}, we focus on the
$O(N_f^{-1})$ corrections, which depend on the actual number of flavor
components.

In our numerical simulations we use the KS staggered formulation of
Ref.~\cite{HKK-92}, with scalar fields located at the sites of the
dual lattice. To compare with the large-$N_f$ predictions obtained for
the GN QFT (\ref{lagrangianGN}), we perform simulations for
$N_f=4,8,12,16$.  We anticipate that our numerical results for the
critical behavior of lattice KS formulations of GN models agree with
the available large-$N_f$ QFT results, thus supporting the conjectured
realization of the GN QFTs through the critical behavior of lattice GN
models.

The paper is organized as follows. In Sec.~\ref{models} we present the
lattice KS formulation of the GN model that we consider.
Sec.~\ref{FSS} outlines the strategy of our analysis of the numerical
data.  In Sec.~\ref{lnres} we report the large-$N_f$ expansions of the
critical exponents, which are then compared with numerical
results. Sec.~\ref{numres} is devoted to the presentation of the
numerical results for various flavor numbers $N_f=4,8,12,16$. Finally,
in Sec.~\ref{conclu} we summarize and draw our conclusions. In the
appendices we report a discussion of the relation between the
fermionic condensate and the scalar field, some technical details on
the simulations, and a collection of FSS results.

\section{Lattice formulations}
\label{models}

A {\em naive} lattice formulation of the 3D massless GN model
(\ref{lagrangianGN}) can be obtained by discretizing the Lagrangian
density (\ref{lagrangianGN2}) on a cubic lattice. The lattice
Hamiltonian is~\cite{CER-83}
\begin{eqnarray}
  H_{N} = \sum_{{\bm x},\mu,c}
    \bar{\psi}_{\bm x}^c \Big( \gamma_\mu \Delta_\mu 
+ \sigma_{\bm x}\Big) \psi_{\bm x}^c
+ {1\over2 g^2} \sum_{\bm x} \sigma_{\bm x}^2\,,
\label{naiveham}
\end{eqnarray}
where $c=1,...,N_\psi$ and $\mu=1,2,3$. Here, $\psi_{\bm x}^c$ is a
four-component spinor for each value of the index $c$, $\gamma_\mu$
are the four-dimensional $\gamma$ matrices, and $\Delta_\mu \psi_{\bm
  x}^c = (\psi_{{\bm x}+\hat\mu}^c - \psi_{{\bm x}-\hat\mu}^c)/2$.  We
set the lattice spacing $a=1$.
One can easily verify that model (\ref{naiveham}), as the massless GN
model \eqref{lagrangianGN2}, is invariant under two {\em chiral}
${\mathbb Z}_2$ symmetries:
\begin{equation}
\begin{aligned}
  \psi_{\bm x}^c \to \gamma_5\psi_{\bm x}^c\,,\quad \bar\psi_{\bm x}^c
  \to - \bar\psi_{\bm x}^c
\gamma_5\,,\quad \sigma_{\bm x} \to - \sigma_{\bm x}\,, \\
  \psi_{\bm x}^c \to \gamma_4\psi_{\bm x}^c\,,\quad \bar\psi_{\bm x}^c
  \to - \bar\psi_{\bm x}^c
\gamma_4\,,\quad \sigma_{\bm x} \to - \sigma_{\bm x}\,, \\
\end{aligned}
\label{z2symmlat}
\end{equation}
which protect the fermion field against the generation of mass terms.
Moreover the model (\ref{naiveham}) is also invariant under the global
symmetry group ${\rm U}(N_\psi)\otimes {\rm U}(N_\psi)$.

The above model does not have the exact flavor content of the GN field
theory due to the lattice fermion doubling.  Indeed, the
Hamiltonian~(\ref{naiveham}) actually describes $8 N_\psi$ massless
flavors in the formal continuum limit $a\to 0$.  They are associated
with the Fourier-transform components with $k_\mu = 0$ and
$k_\mu=\pi/a$ along each direction.  However, the scalar-field
interaction breaks the symmetry of the doublers and therefore this
lattice formulation does not describe $N_f=8 N_\psi$ identical flavors
as required by the GN model, but $N_f$ massless flavors with complex
self-interactions that do not reproduce the field theory model with
Lagrangian density ~(\ref{lagrangianGN2}).  As put forward in
Refs.~\cite{CER-83,HKK-92}, the problem can be attenuated by defining
the scalar fields on the dual lattice, i.e., at the center of the
lattice cubes, located at
\begin{equation}
  \widetilde{\bm x}={\bm x} + \sum_\mu \hat{\mu}/2\,,
  \label{dualloc}
  \end{equation}
      where $\hat\mu$ are the unit vectors associated with the
lattice directions.  The Hamiltonian (\ref{naiveham}) is replaced by
\begin{eqnarray}
  H_{D} = \sum_{{\bm x}\mu c}
    \bar{\psi}_{\bm x}^c \Big( \gamma_\mu \Delta_\mu +
{1\over 8}  
\sum_{\langle {\bm x}, \widetilde{\bm x} \rangle} \sigma_{\widetilde{\bm x}} \Big)
\psi_{\bm x}^c  
+ {1\over 2g^2} \sum_{\widetilde{\bm x}} \sigma_{\widetilde{\bm x}}^2\,,\;\;
\label{naiveham2}
\end{eqnarray}
where the second term includes a sum over the eight sites ${\langle
  {\bm x}, \widetilde{\bm x} \rangle}$ of the dual lattice surrounding
${\bm x}$.  Following Ref.~\cite{CER-83}, one can show that the global
flavor symmetry for all $N_f=8 N_\psi$ flavors is recovered at a
classical tree-order level: in the formal expansion of the Hamiltonian
in powers of $a$, the terms breaking the flavor symmetry are of order
$a$ in the $a\to 0$ limit.  Therefore, at least in the classical
limit, apart from $O(a)$ corrections, one effectively recovers the
${\rm U}(N_f)\otimes {\rm U}(N_f)$ symmetry with $N_f=8 N_\psi$. The
Hamiltonian~\eqref{naiveham2} is also exactly invariant under chiral
${\mathbb Z}_2$ symmetry transformations.  They are defined as in
Eq.~(\ref{z2symmlat}) with $\sigma_{\bm x}$ replaced by
$\sigma_{\widetilde{\bm x}}$.

To reduce the problem of fermionic doubling, one can consider the
staggered KS formulation.  For 3D systems defined on cubic lattices,
there are only two doublers instead of the eight ones appearing in the
naive formulation.  By using $N_\chi$ staggered fermionic fields
$\chi_{\bm x}^c$, we obtain a lattice formulation with $N_f=2N_\chi$
effective massless flavors. The partition function is~\cite{HKK-92}
\begin{eqnarray}
Z &=& \int [\dd{\chi}\dd{\bar{\chi}}][\dd{\sigma}]
  e^{-H_S[\bar{\chi}, \chi, \sigma]}\,, \label{partition}\\
H_{S} &=& \sum_{{\bm x}, {\bm y}, c}
  \bar{\chi}^c_{\bm x}
  M_{{\bm x},{\bm y}} \chi^c_{\bm y} +
\frac{\kappa N_\chi}{2}\sum_{\widetilde{\bm x}}\sigma^2_{\widetilde{\bm x}}\,,
\label{KSmod}
\end{eqnarray}
where $\chi_{\bm x}^c$ is defined on the sites ${\bm x}$ of the cubic
lattice (the index $c$ runs from 1 to $N_\chi$), $\sigma_{\widetilde{\bm x}}$ is
a real scalar field on the dual lattice site $\widetilde{\bm x}$, and
$\kappa$ is the model parameter that is tuned to approach the critical
point~\cite{HKK-92}. The matrix $M$ is given by
\begin{equation}
  M_{{\bm x},{\bm y}}(\sigma) = \sum_{\mu=1}^d \frac{\eta_\mu(\bm x)}{2}
  (\delta_{{\bm y},{\bm x}+\hat\mu} - \delta_{{\bm y},{\bm
        x}-\hat\mu}) + \frac{1}{8} \sum_{\langle {\bm x}, \widetilde{\bm x} \rangle}
        \sigma_{\widetilde{\bm x}} \delta_{{\bm x},{\bm y}}\,.
\label{mmatrix}
\end{equation}
In the above expression, the second sum is over the sites of the dual
lattice that surround ${\bm x}$, and $\eta_\mu({\bm x})$ is the
Kawamoto-Smit phase $\eta_{\mu}({\bm x})=(-1)^{x_1+..+x_{\mu-1}}$.
The matrix $M_{{\bm x}, {\bm y}}$ satisfies the relation
\begin{equation}
  M_{{\bm x}, {\bm y}}(-\sigma) = - M_{{\bm y}, {\bm x}}(\sigma)\,.
  \label{def_M_minus_sigma}
\end{equation}
Apart from irrelevant normalization constants, by integrating out the
fermionic variables we obtain the partition function
\begin{eqnarray}
  Z = \int [\dd{\sigma}]\det \big[M_{{\bm x}, {\bm y}}(\sigma) \big]^{N_\chi}
  \exp(-\frac{\kappa N_\chi}{2}\sum_{\widetilde{\bm x}}
  \sigma^2_{\widetilde{\bm x}})\,. \label{partition_only_sigma}
\end{eqnarray}
The staggered KS formulation (\ref{KSmod}) maintains an exact {\em
  chiral} ${\mathbb Z}_2$ symmetry, corresponding to the
  \begin{equation}
    \chi^c_{\bm x}\to P_{\bm x}\,\chi^c_{\bm x}\,,\;\;\; \bar{\chi}^c_{\bm x}
    \to  -P_{\bm x} \,\bar{\chi}^c_{\bm x}\,,
    \;\;\; \sigma_{\widetilde{\bm x}}\to-\sigma_{\widetilde{\bm x}}\, ,
  \end{equation}
where $P_{\bm x}=(-1)^{\sum_i x_i}$ is the parity of the site ${\bm x}$. 

As in the naive fermion formulation, the continuum flavor symmetry
${\rm U}(N_f)\otimes{\rm U}(N_f)$ of the continuum GN field theory is
not exact. The exact flavor symmetry group of the lattice model is
only ${\rm U}(N_\chi)$, where $N_\chi=N_f/2$.  However, as shown in
Ref.~\cite{HKK-92}, in the formulation (\ref{KSmod}) with scalar
fields on the dual lattice, the symmetries of the continuum GN field
theory are recovered in the formal classical limit $a\to
0$. Violations are of order $a$ and vanish in the formal continuum
limit.

\section{Finite-size scaling at the chiral transition}
\label{FSS}

We investigate numerically the critical behavior of the lattice
KS formulation, using FSS methods applied to several observables
defined in terms of the scalar and fermionic fields.  In our work
boundary conditions (BC) have been chosen as follows.  For fermionic
fields we use antiperiodic BC in one of the directions (we have chosen
the third direction, $\mu=3$) and periodic BC in the other ones. For
the scalar field $\sigma_{\widetilde{\bm x}}$ we use periodic BC in all
directions.

\subsection{Observables}
\label{obs}

We define the two-point function of the $\sigma_{\widetilde{\bm x}}$
field as
\begin{equation}
  G_\sigma({\widetilde{\bm x}}- {\widetilde{\bm y}}) =
  \langle \sigma_{\widetilde{\bm x}} \sigma_{\widetilde{\bm y}}\rangle\,
  \label{twosca}
\end{equation}
(which only depends on ${\widetilde{\bm x}}- {\widetilde{\bm y}}$
because of the translation invariance preserved by the periodic BC)
and the corresponding Fourier transform $\widetilde{G}_\sigma({\bm
  p})=\sum_{\widetilde{\bm x}} e^{i {\bm p} \cdot \widetilde{\bm x}}
G_\sigma({\widetilde{\bm x}})$.  The scalar susceptibility $\chi_\sigma$
and second-moment correlation length $\xi_\sigma$ are defined as
\begin{eqnarray}
  &&  \chi_\sigma = \widetilde{G}_\sigma(0)\,, \label{defchi}\\
  &&\xi_\sigma^2 = \frac{1}{4 \sin^2({{\bm
        p}_m}/2)}\frac{\widetilde{G}_\sigma({\bm 0}) - \widetilde{G}_\sigma({{\bm
      p}_m})}{\widetilde{G}_\sigma({{\bm p}_m})}\,,
\label{defxi}
\end{eqnarray}
where ${\bm p}_m\equiv(0, 0, 2\pi/L)$ (the $z$ direction is the one in
which we use antiperiodic BC for the fermionic variables).  We also
consider the space average of the scalar order parameter
\begin{equation}
\Sigma = \frac{1}{V}\sum_{\widetilde{\bm x}}\sigma_{\widetilde{\bm x}}\,
\end{equation}
(note that $\langle\Sigma\rangle=0$ 
because of the chiral symmetry), and the corresponding 
Binder parameters
\begin{eqnarray}
U_4 = \frac{\expval{\Sigma^4}}{\expval{\Sigma^2}^2}\,,\qquad
U_2 = 
\frac{\expval{\Sigma^2}}{\expval{|\Sigma|}^2}\,.\quad
  \label{def_Us}
  \end{eqnarray}
The observables $U_4$, $U_2$, and $R_\xi\equiv \xi_\sigma/L$, are RG
invariant at the transition where the scalar-field and fermionic
correlations are critical.  They will play a central role in our
numerical FSS analyses.

We define the fermionic susceptibility $\chi_\chi$ as
\begin{equation}
  \chi_\chi \equiv \frac{1}{V}\big\langle\big| \sum_{{\bm x},{\bm
      y}}\bar{\chi}^c_{\bm x}\chi^c_{\bm y}\big|\big\rangle =
  \frac{1}{V}\big\langle\big|\sum_{{\bm x}, {\bm y}}M^{-1}_{{\bm x},
    {\bm y}}\big|\big\rangle\,.
    \label{def_fermioni_susc}
\end{equation}
The absolute value in Eq.~\eqref{def_fermioni_susc} is required by the
presence of the $\mathbb{Z}_2$ invariance. Indeed, using
Eq.~\eqref{def_M_minus_sigma} one can easily prove that
$\big\langle\sum_{{\bm x}, {\bm y}}M^{-1}_{{\bm x},{\bm
    y}}\big\rangle$ vanishes.

We also define the fermionic bilinear $\Xi_{\bm x} = \sum_c
\bar{\chi}_{\bm x}^c \chi_{\bm x}^c$ and its space average
\begin{equation}
  \Xi = {1\over V} \sum_c \sum_{\bm x} \bar{\chi}_{\bm x}^c \chi_{\bm x}^c\,.
\end{equation}
Because of the chiral ${\mathbb Z}_2$ symmetry
\begin{equation}
\expval \Xi =
    \frac{1}{V}\langle\Tr M^{-1}\rangle = 0\, .
    \label{def_condensate}
\end{equation}
The average value of higher powers of $\Xi$ can be directly related to
averages of the scalar order parameter $\Sigma$. Indeed, in the
infinite-volume limit, see Appendix \ref{App.A}, we have
\begin{equation}
\langle \Xi^{n} \rangle = \kappa^n N_\chi^n \langle \Sigma^n\rangle\,.
\label{EQXiN}
\end{equation}
More generally, see Appendix~\ref{App.A}, correlations of the $\sigma$
and of the $\bar{\chi} \chi$ operator are directly related, apart from
contact terms.  This relation expresses the equivalence of the
$\sigma$ and of the $\bar{\chi} \chi$ operator. The presence of
contact terms is not unexpected, because of the different nature of
the two quantities.  For instance, $(\bar{\chi}\chi)^{N_\chi}$
vanishes because of the Grassmannian nature of the variables, while
obviously $\sigma^{N_\chi}$ is nonvanishing.

\subsection{FSS strategy to determine the critical exponents}
\label{fssstrategy}

We now present the FSS relations we will use in the numerical
analysis.  To estimate the correlation-length exponent $\nu$ and the
critical value $\kappa_c$, we analyze the behavior of RG invariant
quantities $R$ (such as $U_4$, $U_2$, and $R_\xi \equiv \xi_\sigma/L$,
defined in Sec.~\ref{obs}). Close to the critical point $\kappa =
\kappa_c$, they are expected to behave as
\begin{eqnarray}
  R(\kappa,L) \approx {\cal R}(X)\,,\qquad X =
  (\kappa-\kappa_c)L^{1/\nu}\,.
  \label{rscal}
\end{eqnarray}
The function ${\cal R}(X)$ is universal up to a multiplicative
rescaling of its argument.  In particular, $R^*\equiv {\cal R}(0)$ is
universal, depending only on the boundary conditions and aspect ratio
of the lattice.  Eq.~(\ref{rscal}) holds up to scaling corrections,
decaying as $L^{-\omega_l}$, where $\omega_l>0$ is the leading
scaling-correction exponent.

If a RG invariant quantity $\widehat{R}$ is a monotonic function of
$X$---this is the case of the ratio $R_\xi=\xi_\sigma/L$---in the FSS
limit we can express a different RG invariant quantity $R$ as a
function of $\hat{R}$ simply as
\begin{equation}
R(\kappa,L) = F_R(\widehat{R}) + O(L^{-\omega_l})\,,
\label{rscal2}
\end{equation}
where $F_R(x)$ depends only on the universality class, boundary
conditions and lattice shape, without nonuniversal multiplicative
factors.  Scaling \eqref{rscal2} is particularly convenient to test
universality-class predictions, since it permits easy comparisons
between different models without requiring the tuning of nonuniversal
parameters.

Another independent critical exponent is $\eta_\sigma$. It can be
defined in terms of the critical behavior of the two-point function
$G_\sigma$ defined in Eq.~(\ref{twosca}). In the thermodynamic limit
and at the critical point, $G_\sigma({\bm z})$ behaves as
\begin{equation}
G_\sigma({\bm z}) \sim |{\bm z}|^{-2y_\sigma}, 
\label{gsigcrit}
\end{equation}
where $y_\sigma=(d-2+\eta_\sigma)/2$ is the RG dimension of the scalar
field $\sigma$.  The exponent $\eta_\sigma$ can be estimated from the
FSS behavior of the scalar susceptibility $\chi_\sigma$ defined in
Eq.~(\ref{defchi}), which is expected to scale as
\begin{equation}
  \chi_\sigma(\kappa,L) \approx  L^{2-\eta_\sigma} {\cal S}(X)\,,
\label{eta-RG1}
\end{equation}
where ${\cal S}$ is a universal function apart from an overall factor
and a rescaling of the argument. We can also replace $X$ with a
monotonic RG invariant quantity $\widehat{R}$, as
\begin{equation}
  \chi_\sigma(\kappa,L) \approx L^{2-\eta_\sigma}
  F_\sigma(\widehat{R})\,,
\label{eta-RG2}
\end{equation}
where $F_\chi(\widehat{R})$ is universal apart from a multiplicative
factor only.  The critical exponent $\eta_f$, related to the RG
dimension of the fermionic field $y_\psi = (d-1+\eta_f)/2$, can be
obtained from the analysis of the fermionic susceptibility $\chi_\chi$
defined in Eq.~(\ref{def_fermioni_susc}). In the FSS limit, it
satisfies the scaling relation
\begin{equation}
  \chi_\chi(\kappa,L) \approx L^{1-\eta_f}
  F_\chi(\widehat{R})\,.
  \label{etaf-RG}
\end{equation}

\section{Large-$N_f$ results within QFT}
\label{lnres}

In this section, we report the known leading terms of the large-$N_f$
expansion of the exponents $\nu$, $\eta_\sigma$, and $\eta_f$ defined
in the previous section.  They are given
by~\cite{Gracey-93,Gracey-94,MZ-03,EIKLPS-22}
\begin{eqnarray}
 {1\over \nu} &\approx& 1 - {8\over 3 \pi^2} N_f^{-1}
  + {4(27\pi^2+632)\over 27\pi^4} N_f^{-2} \,,\label{invnuln}\\
  \eta_\sigma &\approx& 1 - {16\over 3 \pi^2} N_f^{-1}
  -  {4(27\pi^2-304)\over 27\pi^4} N_f^{-2} \,,\label{etaphiln}\\
    \eta_f &\approx& {2\over 3 \pi^2} N_f^{-1}
  + {122 \over 27\pi^4} N_f^{-2}  \label{etapsiln}\\
  &+&   {4\over 27\pi^6}\left({47\pi^2\over 12} + 9 \pi^2\ln 2
  - {189\over 2}\zeta(3)- {167\over 9}\right) N_f^{-3}\,.
  \nonumber
\end{eqnarray}
The $1/N_f$ expansion allows us to predict also the exponents of the
scaling corrections. First, there are scaling corrections related to
the irrelevant fields that appear in the continuum GN QFT. The
correspondent leading scaling-correction exponent is~\cite{Gracey-17}
\begin{eqnarray}
\omega = 1 - {32\over 3 \pi^2} N_f^{-1}
  \,.\label{omegtaln}
\end{eqnarray}
There are also scaling corrections that are specific of the lattice
model.  The most relevant ones are associated with the operator that
breaks the flavor symmetry. Since this term is formally of order $a$
in the continuum tree-level approximation~\cite{HKK-92}, we can
predict that the corresponding correction-to-scaling exponent
$\omega_d$ is 1 for $N\to\infty$, i.e.,
\begin{equation}
  \omega_d = 1 + O(N_f^{-1})\,.
  \label{omegadest}
\end{equation}
The exponents defined in Eqs.~(\ref{omegtaln}) and (\ref{omegadest})
coincide for $N_f=\infty$. For finite values of $N_f$, they differ (we
do not know which one is the smallest), but, if $N_f$ is large, they
should be still close enough to justify the use of a single correction
to scaling with exponent $\omega \approx 1$.  For small values of
$N_f$ their difference might be significant. In this case, the
presence of two different correction terms would make the numerical
analysis quite challenging.

We finally mention that the critical value $\kappa_c$ for the KS
formulation (\ref{KSmod}) was computed in the limit $N_\chi\to\infty$,
obtaining~\cite{HKK-92,GZ-77,JZ-01}
\begin{equation}
  \kappa_{c,\infty} = 2 \int_0^\infty dz \, e^{-3z} I_0^3(z) =
{(\sqrt{3}-1) \Gamma(\frac{1}{24})^2 \Gamma(\frac{11}{24})^2 \over 48 \pi^3} 
  \label{wcinfty}
\end{equation}
where $I_0$ is the modified Bessel function. Numerically, we find
$\kappa_{c,\infty}=1.010924039\ldots$ 

\section{Numerical analyses} 
\label{numres}

We now outline our numerical FSS analyses of the MC data. We simulate
the staggered KS model (\ref{KSmod}) using the hybrid MC algorithm,
see also App.~\ref{app_montecarlo} for technical details.  We present
results for various numbers of massless flavors, i.e., for
$N_f=4,8,12,16$, to check the approach to the large-$N_f$ limit. They
correspond to $N_\chi=2,4,6,8$ equal staggered components $\chi_{\bm
  x}^c$.  Even numbers of $N_\chi$ are required to avoid the sign
problem in the MC simulations, see also App.~\ref{app_montecarlo}.

The efficiency of the hybrid MC algorithms for fermionic models
significantly decreases when increasing the size of the lattice:
autocorrelation times generally increase with a large power of $L$
\cite{Kennedy-99}. The computational cost in our MC simulations
appears to approximately increase as $L^6$ in the critical region, see
App.~\ref{app_montecarlo} for some details.  For this reason, we
performed simulations on relatively small lattices, up to $L\approx
40$, where it was possible to obtain accurate data. Obtaining precise
estimates for larger lattice sizes would require a much larger
numerical effort.

\begin{figure}
    \centering \includegraphics[width=0.95\columnwidth]{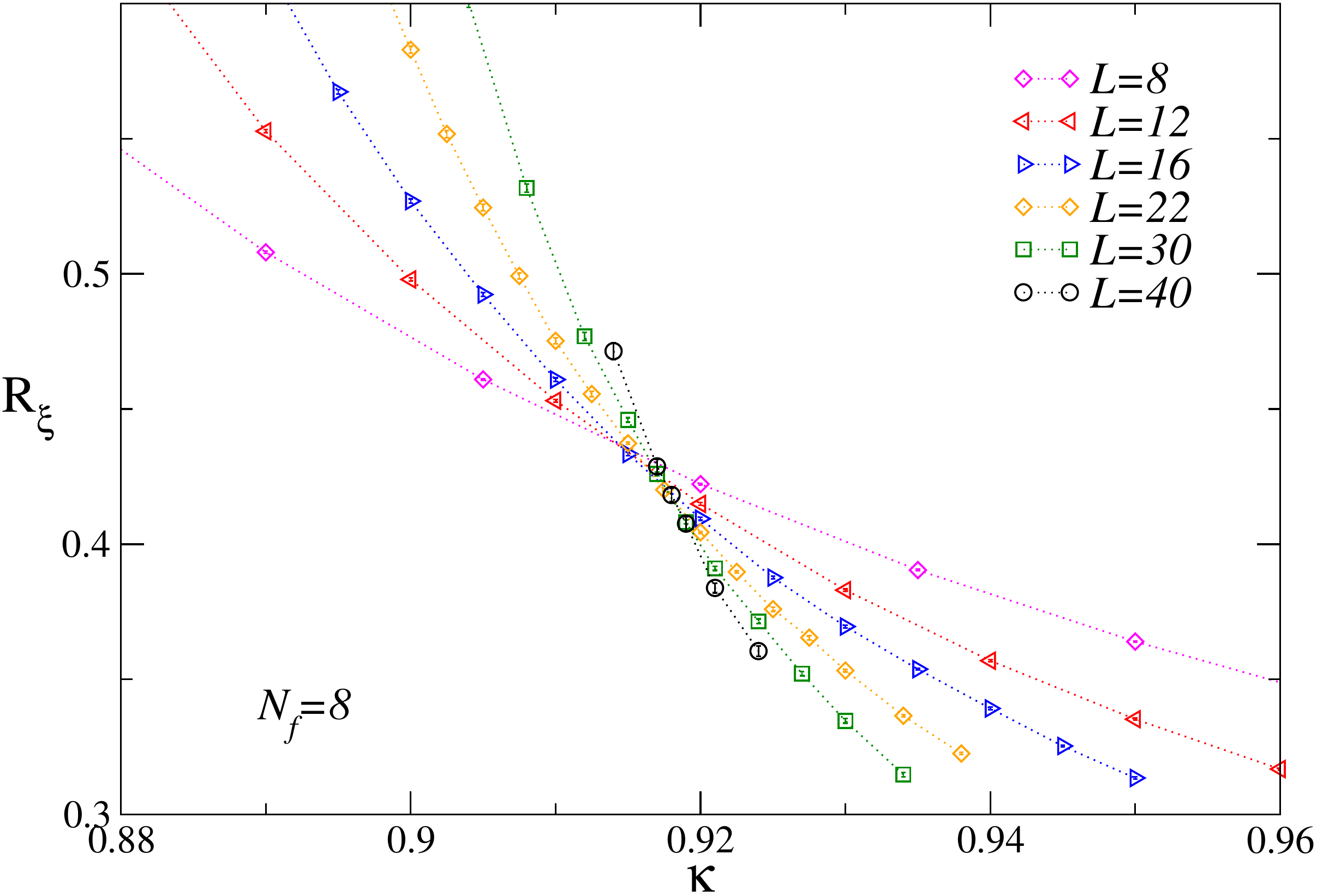}
    \caption{MC estimates of $R_\xi$ versus $\kappa$ for $N_f=8$.  The
      data for for different lattice sizes have a crossing point for
      $\kappa_c\approx 0.92$.}
    \label{rxinf8}
\end{figure}

The FSS analysis of the MC data shows clear evidence of a continuous
chiral transition for all values of $N_f$ considered.  The MC
estimates of the RG invariant quantities $R_\xi$, $U_4$, $U_2$ defined
Sec.~\ref{obs}, show a clear crossing point; see, e.g.,
Fig.~\ref{rxinf8}, where we report $R_\xi$ as a function of $\kappa$
for $N_f=8$.

To determine the critical point $\kappa_c$ and the exponent $\nu$, we
fitted $R_\xi$, $U_4$, and $U_2$ to the general FSS relation
(\ref{rscal}). We performed fits parametrizing ${\cal R}(X)$ with a
poynomial in $X$, including only data satisfying $L\ge L_{\rm min}$,
to identify scaling corrections. We also performed combined fits of
pairs of observables to
\begin{equation}
  R(\kappa,L) = {\cal R}(X) + L^{-\omega_l} {\cal R}_c(X),
\end{equation}
fixing $\omega_l = 1$ (this should be a reasonable estimate for $N_f$
large, as discussed in Sec.~\ref{lnres}). The results show some tiny
trends both for $\kappa_c$ and $\nu$ and also some dependence on the
observable considered.  Scaling corrections, numerically large
compared to our tiny error bars, are clearly present. As an example,
we report the estimates of $\kappa_c$ for $N_f=12$, obtained from the
analysis of the data in the range $-0.3\le X\le 0.3$.  The analysis of
$R_\xi$ provides $\kappa_c = 0.9463(1), 0.9470(1)$, for $L_{\rm
  min}=12$ and 16, respectively. The analysis of $U_4$ gives instead
$\kappa_c = 0.9472(1), 0.9479(3)$. It is clear that the statistical
error is negligible compared with the systematic error due to the
scaling corrections.  If we consider the differences of these numbers
as an estimate of the systematic uncertainty, we end up with
$\kappa_c=0.9627(7)\,,0.9475(6),\,0.9180(5),\,0.8348(8)$ for
$N_f=16,12,8,4$, respectively. Our result for $N_f=4$ is in perfect
agreement with the estimate reported in Ref.~\cite{CS-07},
$\kappa_c=0.835(1)$.

\begin{figure}
    \centering
    \includegraphics[width=0.95\columnwidth]{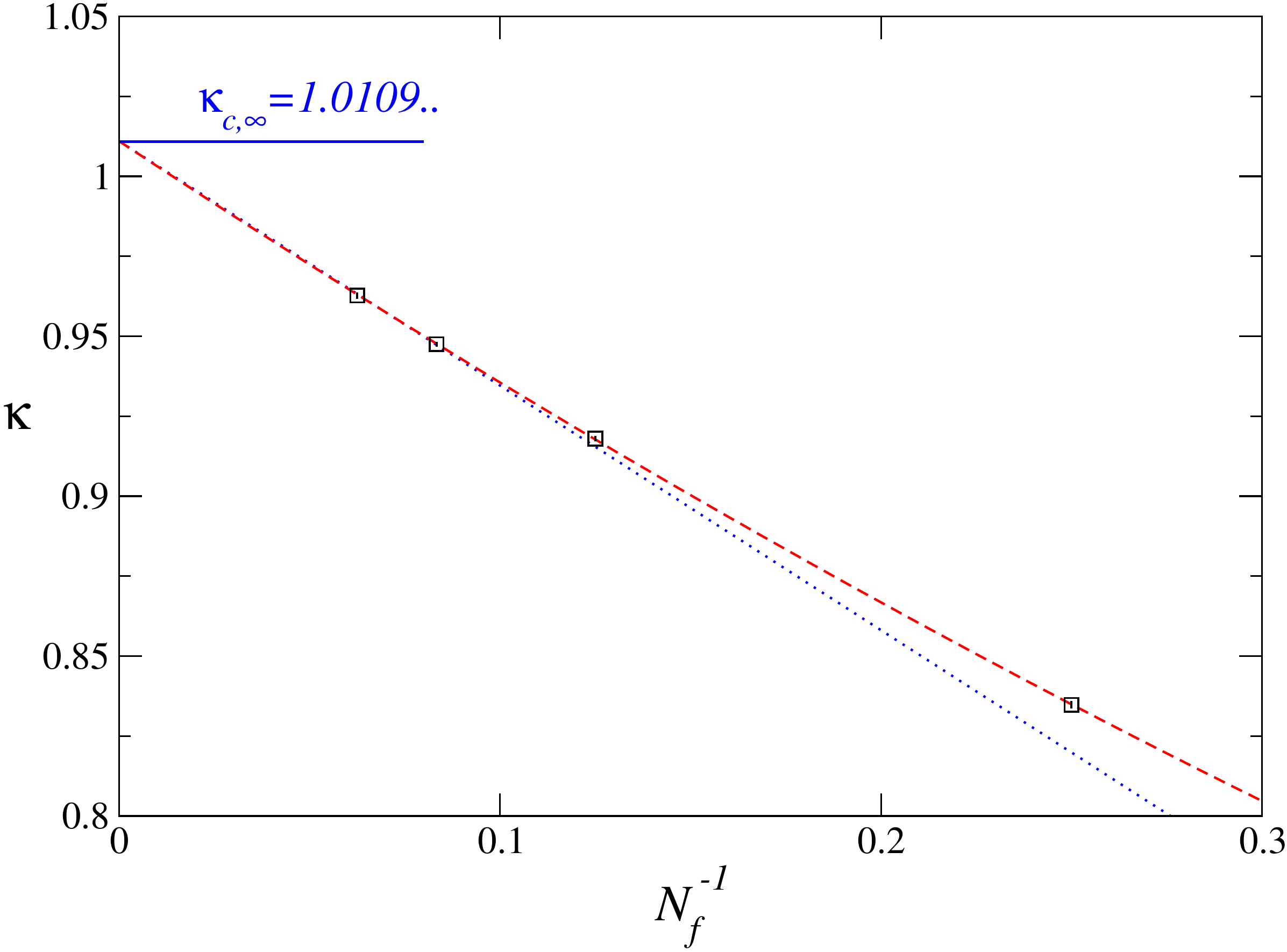}
    \caption{Estimates of the critical value $\kappa_c$ versus
      $N_f^{-1}$. They are fully consistent with the behavior
      $\kappa_c(N_f) = \kappa_{c,\infty} + a_1 N_f^{-1}$, where
      $\kappa_{c,\infty}\approx 1.0109$ is the exact result for
      $N_f=\infty$, see Eq.~(\ref{largenapp}).  The blue dotted line
      corresponds to $\kappa_c(N_f) = \kappa_{c,\infty} + a_1
      N_f^{-1}$; a fit of the data with $N_f=12$ and 16 gives $a_1 =
      -0.764(4)$.  The red dashed line corresponds to $\kappa_c(N_f) =
      \kappa_{c,\infty} + a_1 N_f^{-1} + a_2 N_f^{-2}$, where the
      coefficients were obtained by fitting all data ($\chi^2/{\rm
        d.o.f}\approx 0.4$): $a_1=-0.788(4)$ and $a_2=0.33(2)$.  }
    \label{figbetaclargeNf}
\end{figure}

The estimates of $\kappa_c$ appear to approach the $N_f\to\infty$
critical value $\kappa_{c,\infty}\approx 1.0109$ with increasing
$N_f$, cf.  Eq.~(\ref{wcinfty}), as shown in
Fig.~\ref{figbetaclargeNf}. Actually, they appear to converge to the
$N_f=\infty$ critical value as
\begin{equation}
  \kappa_c(N_f) = \kappa_{c,\infty} + a_1 N_f^{-1} + O(N_f^{-2})
\label{largenapp}
\end{equation}
with $a_1\approx -0.8$.
  
\begin{figure}
    \centering \includegraphics[width=0.95\columnwidth]{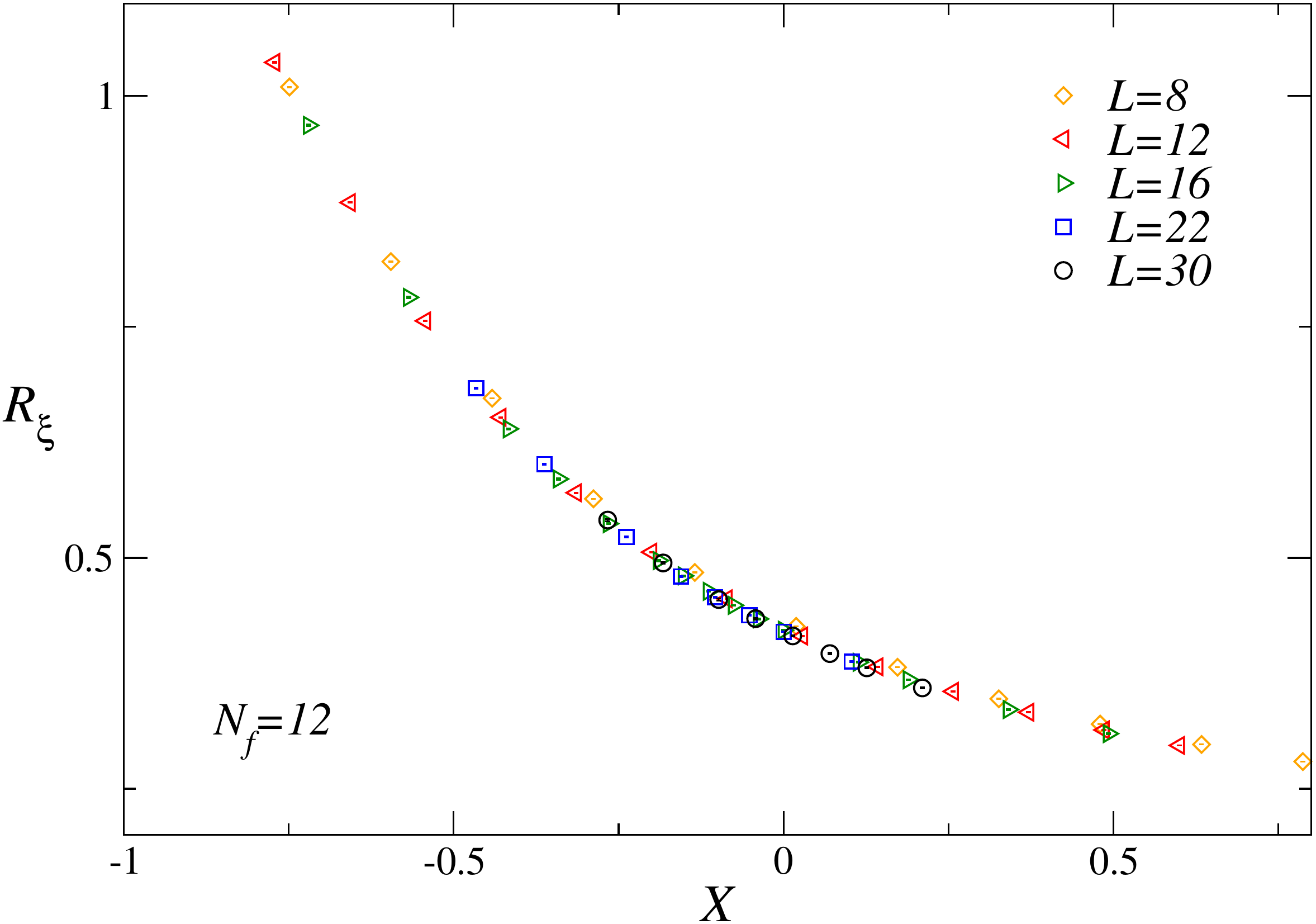}
    \caption{Plot of $R_\xi$ versus $X=(\kappa-\kappa_c)L^{1/\nu}$ for
      $N_f=12$, using the MC estimates $\kappa_c=0.9475$ and
      $\nu=1.02$. }
    \label{rxiscaX}
\end{figure}

The same fits that determine $\kappa_c$ provide estimates of the
critical exponent $\nu$. They are reported in Table~\ref{tableexp}.
Again the error takes into account the small differences obtained from
the analyses of $R_\xi$, $U_4$, and $U_2$. Also for $\nu$, the
differences among the estimates obtained by analyzing the differerent
observables are larger than the statistical error of the fits,
indicating the presence of scaling corrections somehow larger than the
statistical errors. As an example we report the results for $N_f=12$
(the corresponding estimates of $\kappa_c$ are reported above). For
$L_{\rm min} = 12,16$ we obtain $\nu = 1.027(5)$, 1.04(1) from the
analysis of $R_\xi$, and $\nu = 1.01(1)$, 1.00(1), from the analysis
of $U_4$, which are somewhat inconsistent at the level of the
(relatively small) statistical errors.  The final estimates of $\nu$
are in good agreement with the estimates (column LN in
Table~\ref{tableexp}) obtained by using the large-$N_f$ expansion
(\ref{invnuln}) to order $N_f^{-2}$.  As an example of the quality of
the observed scaling, in Fig.~\ref{rxiscaX} we plot $R_\xi$ versus
$X=(\kappa-\kappa_c)L^{1/\nu}$ for $N_f=8$. On the scale of the
figure, all data fall on top of a single curve. Similar plots are
obtained for the other values of $N_f$.

\begin{table}
\begin{tabular}{ccccccc}
\hline\hline
\multicolumn{1}{c}{$N_f$} &
\multicolumn{2}{c}{$\nu$} & 
\multicolumn{2}{c}{$\eta_\sigma$} & 
\multicolumn{2}{c}{$\eta_f$} \\\hline
\multicolumn{1}{c}{} &
\multicolumn{1}{c}{$\quad$ MC $\;\;$} &
\multicolumn{1}{c}{$\;\;$ LN $\quad$} &
\multicolumn{1}{c}{$\quad$ MC $\;\;$} &
\multicolumn{1}{c}{$\;\;$ LN $\quad$} &
\multicolumn{1}{c}{$\quad$ MC $\;\;$} &
\multicolumn{1}{c}{$\;\;$ LN $\quad$} \\\hline
16 &  1.00(2) & 1.0118 & 0.94(3) & 0.9664 & 0.00(1) & 0.0044 \\
12 &  1.02(2) & 1.0135 & 0.92(2) & 0.9554 & 0.01(1) & 0.0059 \\
8 &  1.00(2) & 1.0136 & 0.90(3) & 0.9333 & 0.01(1) & 0.0092 \\
4 &  0.99(1) & 0.9867 & 0.83(2) & 0.8685 & 0.03(2) & 0.0197 \\
\hline\hline
\end{tabular}
\caption{Estimates of the universal critical exponents $\nu$,
  $\eta_\sigma$ and $\eta_f$, obtained in this paper (MC). We also
  report the large-$N_f$ estimates (LN) obtained using the expansions
  Eqs.~(\ref{invnuln}), (\ref{etaphiln}) and (\ref{etapsiln}). For the
  exponent $\nu$ we used the direct expansion to order $N_f^{-2}$,
  obtained by inverting Eq.~(\ref{invnuln}).  }
  \label{tableexp}
\end{table}

\begin{figure}
    \centering \includegraphics[width=0.95\columnwidth]{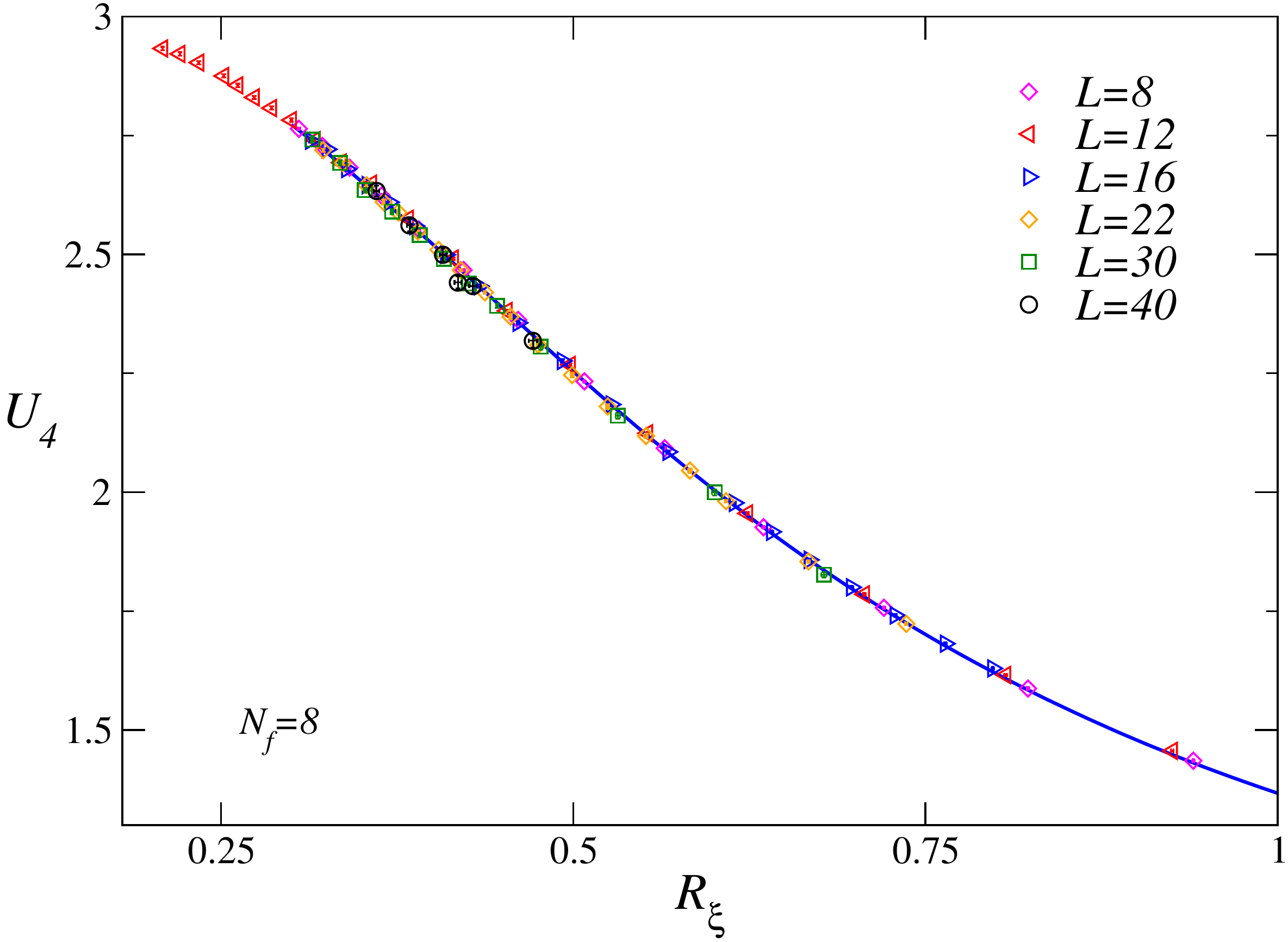}
    \caption{Plot of $U_4$ versus $R_\xi$ for $N_f=8$.  The data
      clearly approach a universal FSS curve, as predicted by the FSS
      Eq.~(\ref{rscal2}).  The blue straight line represents the
      large-size interpolation of the data reported in
      App.~\ref{interpolation}.}
    \label{u4vsrxi}
\end{figure}

\begin{figure}
  \centering
    \includegraphics[width=0.95\columnwidth]{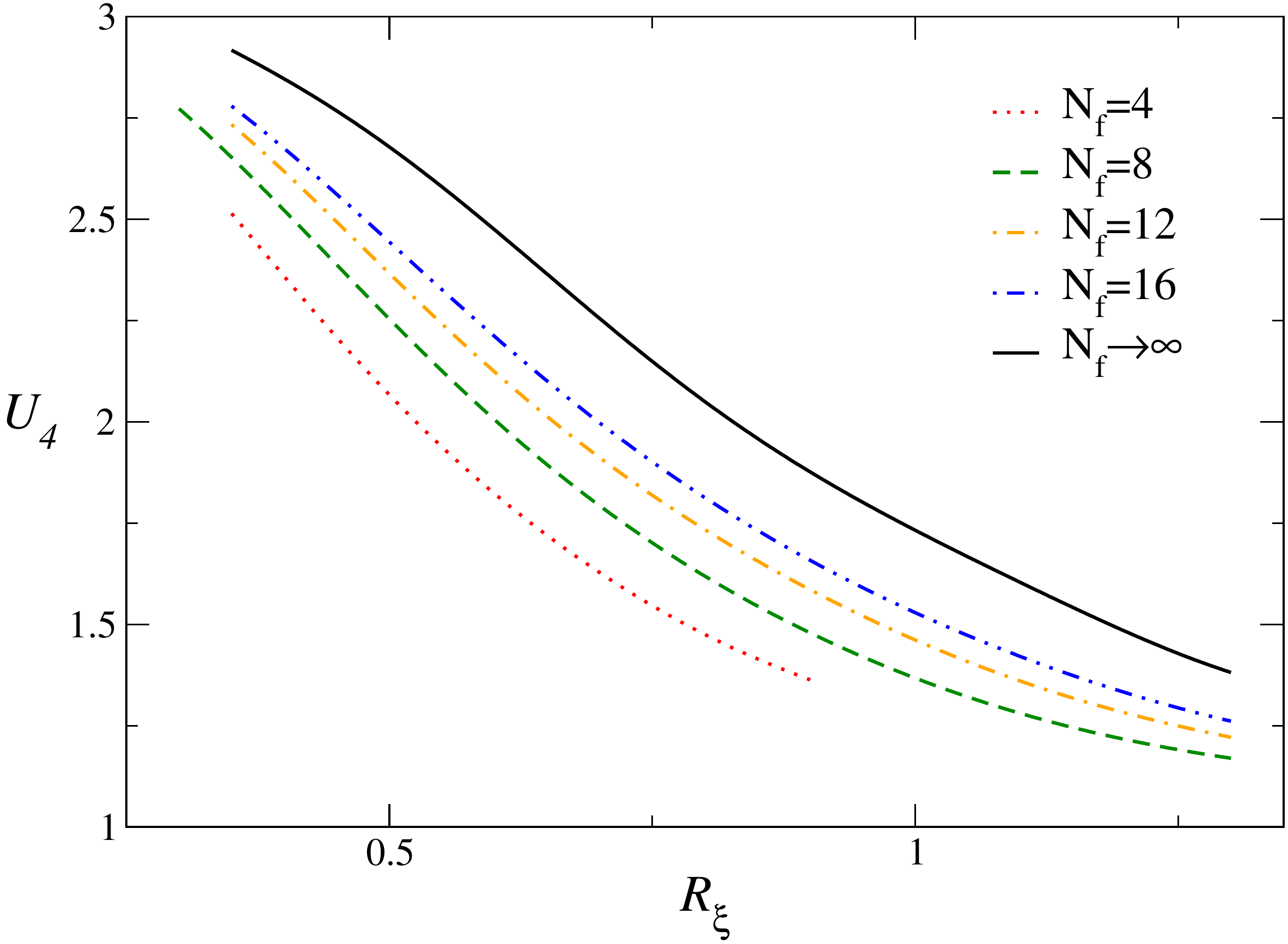}
    \caption{FSS curves of $U_4$ versus $R_\xi$ for $N_f=4,8,12,16$,
      as obtained by interpolating the data for the largest available
      lattices, see App.~\ref{interpolation}. The curves are clearly
      different, confirming that the universality class of the chiral
      transition depends on $N_f$. We also report (black continuous
      line) an estimate of the $N_f=\infty$ curve: it is an
      extrapolation of the results for $N_f=12$ and 16 assuming a
      linear $1/N_f$ approach.}
    \label{u4vsrxialln}
\end{figure}

To obtain a better check of the validity of FSS and verify that
scaling corrections are small, we can use relation ~(\ref{rscal2})
which should hold in the FSS limit, without the need of fixing any
normalization.  As an example, in Fig.~\ref{u4vsrxi} we plot $U_4$
versus $R_\xi$ for $N_f=8$.  The data sets for different values of $L$
approach a universal curve with increasing $L$, as predicted by the
FSS theory.  Scaling corrections are very small on the scale of the
figure. However, at a closer look one observes a systematic downward
drift of the order of the statistical errors on $U_4$ [for $L\le 30$,
  typical errors on $R_\xi$ are smaller than $10^{-3}$, while errors
  on $U_4$ are $O(10^{-3})$].  Analogous plots are obtained for $U_2$,
and for other values of $N_f$. For each $N_f$, the data for $U_4$
versus $R_\xi$ for $L\ge 16$ have been interpolated using
polynomials. These interpolations are reported in
App.~\ref{interpolation} and shown in Fig.~\ref{u4vsrxialln}.  The
curves for different values of $N_f$ clearly differ and appear to
converge to a nontrivial large-$N_f$ curve, which is obtained by
performing an extrapolation assuming a $1/N_f$ correction. The result
of the extrapolation of the curves for $N_f=12$ and 16 is reported in
Fig.~\ref{u4vsrxialln}. An estimate of the error on the extrapolation
can be obtained by considering the extrapolation that uses the data
corresponding to $N_f=8$ and 16, or $N_f=8$ and 12; the resulting
curves differ slightly, the largest deviations are about 3\% and are
observed for $R_\xi\approx 1$.

\begin{figure}
    \centering \includegraphics[width=0.95\columnwidth]{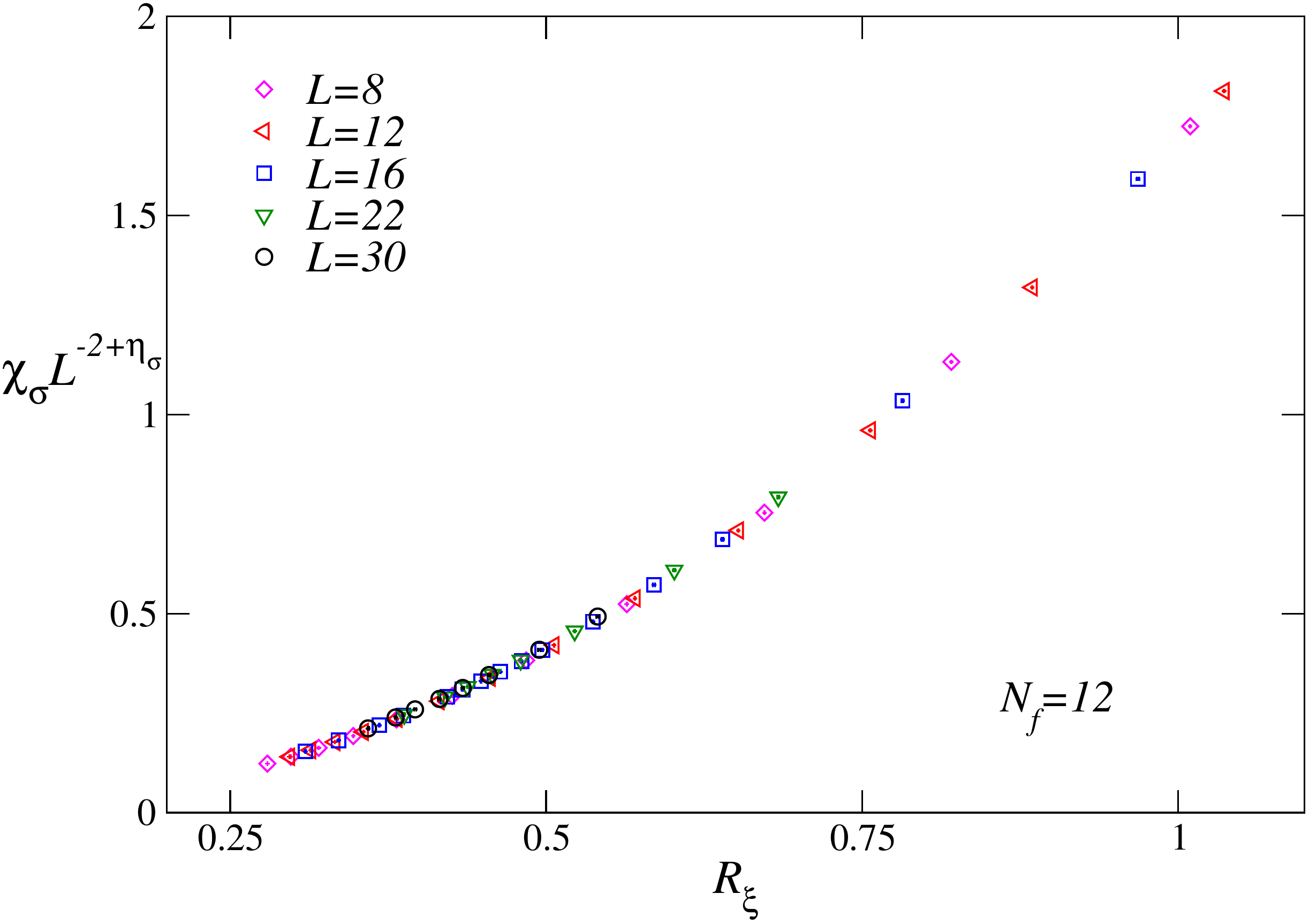}
    \caption{Scaling plot of the scalar susceptibility $\chi_\sigma$
      defined in Eq.~(\ref{defchi}), for $N_f=12$.  We report
      $\chi_\sigma/L^{2-\eta_\sigma}$ versus $R_\xi$. We use the
      estimate $\eta_\sigma=0.92$.  }
    \label{etassca}
\end{figure}

\begin{figure}
    \centering \includegraphics[width=0.95\columnwidth]{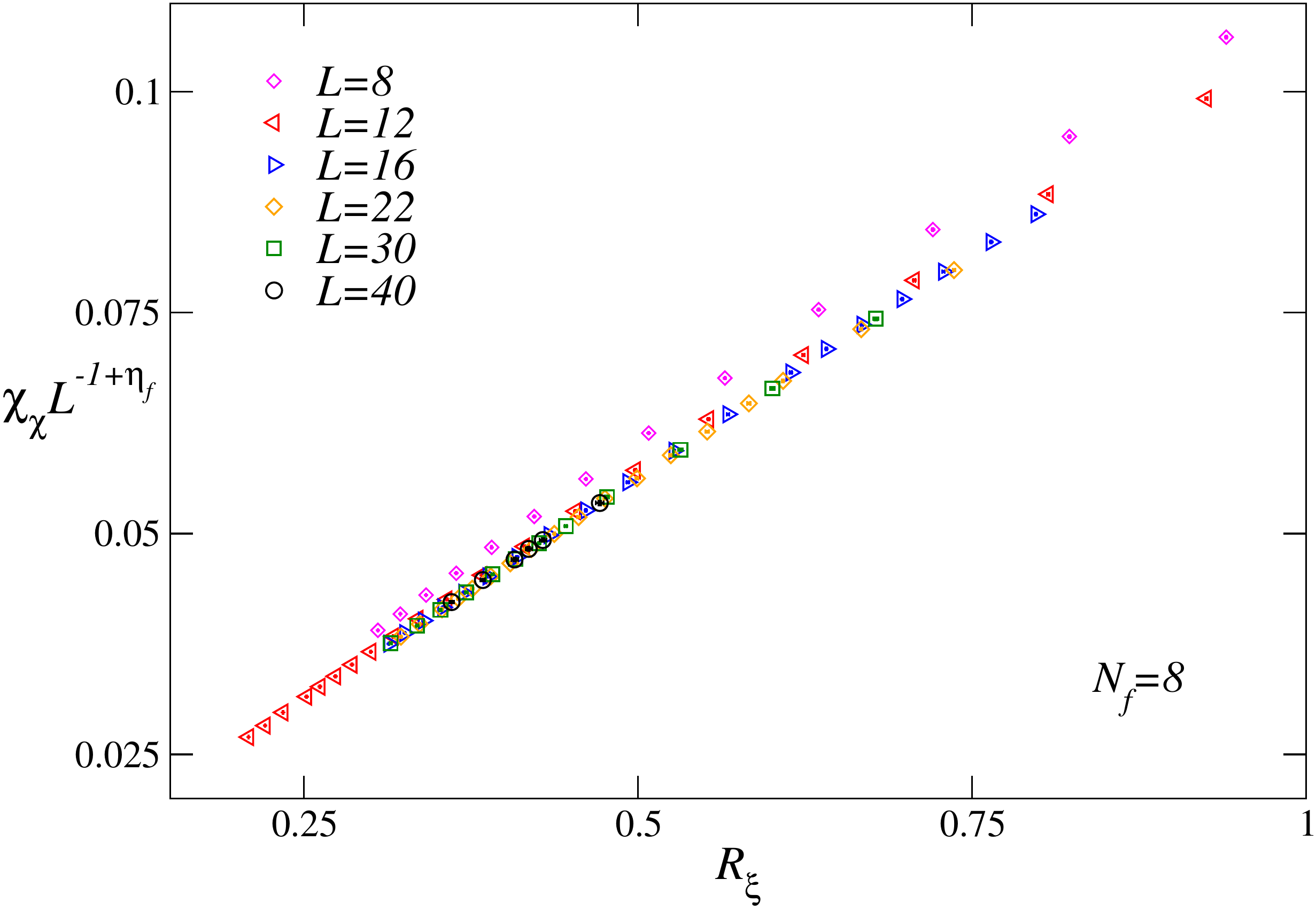}
    \caption{Scaling plot of the fermionic susceptibility $\chi_\chi$
      defined in Eq.~\ref{def_fermioni_susc}), for $N_f=8$.  We report
      $\chi_\chi/L^{1-\eta_f}$ versus $R_\xi$. We use the estimate
      $\eta_f= 0.01$.  }
    \label{etafsca}
\end{figure}

We estimate the critical exponent $\eta_\sigma$ defined in
Eq.~(\ref{gsigcrit}), by analyzing the data for the scalar
susceptibility $\chi_\sigma$. We exploit the FSS relation
(\ref{eta-RG2}) which does not require any knowledge of $\kappa_c$ and
$\nu$, using $R_\xi$, $U_4$ and $U_2$ as arguments. The comparison of
the fit results allows us to estimate the systematic error, which,
again, turns out to be somewhat larger than the statistical error.  In
Table~\ref{tableexp} we report the final estimates. Again we observe a
substantial agreement with the large-$N_f$ estimates obtained using
the expansion~(\ref{etaphiln}).  To show the quality of the scaling of
the scalar susceptibility, in Fig.~\ref{etassca} we report
$\chi_\sigma L^{-2+\eta_\sigma}$ versus $R_\xi$ for $N_f = 12$, using
the final estimate $\eta_\sigma = 0.92$.  On the scale of the figure, we
observe a very good collapse of the data.  Similar plots are obtained
for the other observables and values of $N_f$.

A similar analysis is used to estimate the critical exponent $\eta_f$.
We fit the fermionic susceptibility $\chi_\chi$ to
Eq.~(\ref{etaf-RG}), using $R_\xi$, $U_4$ and $U_2$. In all cases,
fits show a large $\chi^2$ and a systematic drift [the systematic
  deviations are $O(10^{-2})$ and significantly larger than the fit
  statistical error, which is $O(10^{-3})$] towards lower values.  The
errors on the final results, reported in Table~\ref{tableexp}, have
been computed conservatively, looking at all different results
obtained by varying the RG quantity used in the fit and the minimum
value of $L$ of the data that have been considered. Again the final
estimates are consistent with the large-$N_f$ predictions.  In
Fig.~\ref{etafsca} we show the plot of $\chi_\chi L^{-1+\eta_\chi}$
versus $R_\xi$ for $N_f=8$. Scaling corrections are here clearly
visible for $L = 8$.

In conclusion, our numerical estimates of the critical exponents are
in substantial agreement with the large-$N_f$ estimates obtained using
the GN QFT. Thus, they provide a robust evidence that the GN QFT
provides the effective description of the critical behavior of the the
lattice GN model (\ref{KSmod}), with staggered KS fermionic variables
and scalar fields located on the dual lattice. Therefore, the explicit
$O(a)$ breaking of the flavor symmetry occurring in the lattice model
is irrelevant at the chiral critical point, where the global symmetry
enlarges to $U(N_f)\otimes U(N_f)$.

We finally mention that our exponent estimates for $N_f=4$ are in
perfect agreement with those reported in Ref.~\cite{CS-07},
$\nu=0.99(2)$ and $\eta_\sigma=0.835(40)$. Ref.~\cite{CS-07} also
reported the value $U^*_4$ of $U_4$ at the critical point:
$U_4^*=2.304(24)$.  The analysis of our data provides a completely
consistent estimate, $U_4^* = 2.31(1)$. Functional RG results are
reported in Ref.~\cite{RVW-01}.  For both $N_f = 4$ and 12 their
results are consistent with ours.

\section{Conclusions}
\label{conclu}

We present a numerical study of a 3D lattice model with massless
fermions and attractive quartic interactions. We study the critical
behavior at the chiral ${\mathbb Z}_2$ transition to shed light on the
relation between the lattice model and the continuum GN QFT, which is
usually assumed to provide the effective description of the critical
behavior.  In particular, we study the lattice GN model (\ref{KSmod}),
defined in terms of $N_\chi$ staggered KS fermionic variables and of
an auxiliary scalar field located on the dual lattice sites.  The
coupling between the bilinear fermionic operator and the scalar field
is chosen such as to reproduce an attractive quartic interaction among
$N_f = 2 N_\chi$ Dirac fermion fields in the formal continuum limit.
The lattice model is only invariant under global $U(N_\chi)$
transformations.  Thus, the main issue is whether the full flavor
symmetry is recovered in the critical limit, i.e. whether the
long-distance behavior shows an enlarged $U(N_f)\times U(N_f)$
symmetry. In field-theory terms, this would imply that the lattice
operators that break the flavor symmetry are irrelevant in the
critical theory. This is clearly the case for large values of $N_f$
(as we discuss in Sec.~\ref{lnres}, for $N_f=\infty$ the usual formal
argument that these terms are of order $a$, implies that their RG
dimension is $-1$). However, one cannot exclude that they become
relevant for small values of $N_f$.

We present FSS analyses of MC simulations of the lattice GN model
(\ref{KSmod}). We consider massless fermions with $N_\chi$ components,
considering $N_\chi = 2,4,6,8$, which would correspond to
$N_f=4,8,12,16$. A detailed FSS analysis of the numerical data on
lattices of size $L\le 40$ allows us to determine several critical
exponents. We compare the results with those obtained using the GN QFT
with Lagrangian (\ref{lagrangianGN}) in the large-$N_f$ limit, finding
a substantial agreement for all values of $N_f$ considered.  For
$N_f=4$ we also confirm the results of Ref.~\cite{CS-07}.  Our results
confirm that the GN QFT describes the critical behavior of the lattice
GN model (\ref{KSmod}) at the chiral ${\mathbb Z}_2$ transition, even
though the interactions explicitly break the flavor ${\rm
  U}(N_f)\otimes {\rm U}(N_f)$ symmetry of the GN field theory.

The numerical analysis we have presented here indicates that the main
source of error on the estimates of the critical quantities is
systematic. Therefore, to improve the quality of the final results, it
would be crucial to significantly increase the lattice sizes with
comparable accuracy. However, the hybrid MC dynamics shows a strong
critical slowing down, probably also related to the fact that we are
considering the dynamics of a scalar field in a massless fermionic
background. Thus, increasing $L$ requires a large computational
effort.  It is difficult to estimate how large $L$ should be to obtain
a significant improvement, as we have no direct information on the
leading correction-to-scaling exponent $\omega_l$.  For $N_f=\infty$,
we have $\omega_l=1$, but we cannot exclude that $\omega_l$ is
significantly smaller for the values of $N_f$ investigated.

\appendix

\section{Fermionic condensate} \label{App.A}

In this Appendix we derive some relations between correlation
functions of the fermionic condensate and of the scalar field.  To
prove Eq.~(\ref{EQXiN}), we start from the average value of a function
of the $\sigma$ and of the fermionic variables:
\begin{equation}
\langle f \rangle = 
 {1\over Z} \int [d\chi][d\bar{\chi}][d\sigma] 
    e^{-H_s(\bar{\chi},\chi,\sigma)} f(\sigma,\chi,\bar{\chi})\,.
\label{AppA-start}
\end{equation}
Then, we perform the following change of variables:
\begin{equation}
\sigma_{\widetilde{\bm x}} \to   \sigma_{\widetilde{\bm x}}' =
  \sigma_{\widetilde{\bm x}} + \delta_{\widetilde{\bm x},\widetilde{\bm z}}
\epsilon \,,
\end{equation}
where $\widetilde{\bm z}$ is a dual lattice point. Obviously, the
integral appearing in Eq.~(\ref{AppA-start}) is invariant under the
change of variables.  If we write $f(\sigma',\chi,\bar{\chi}) =
f(\sigma,\chi,\bar{\chi}) + \delta_{\widetilde{\bm z}} f \epsilon$ we
obtain the identity
\begin{equation}
- {1\over 8} \sum_{\langle{\bm x},\widetilde{\bm z}\rangle} \sum_c
       \langle \bar{\chi}_{\bm x}^c \chi_{\bm x}^c f\rangle -
    \kappa N_\chi \langle \sigma_{\widetilde{\bm z}} f\rangle +
      \langle \delta_{\widetilde{\bm z}} f \rangle = 0\,.
\label{AppA-relz}
\end{equation}
If we sum over $\widetilde{\bm z}$ and define $\delta f = {1\over V}
\sum_{\widetilde{\bm z}} \delta_{\widetilde{\bm z}} f$, we obtain
\begin{equation}
\langle \Xi f\rangle -  \kappa N_\chi \langle \Sigma f\rangle
 + \langle \delta f \rangle = 0\,.
\end{equation}
Now we use $f(\sigma,\chi,\bar{\chi}) = \Xi^n \Sigma^m$, obtaining
\begin{equation}
   \langle \Xi^{n+1} \Sigma^m \rangle =
        \kappa N_\chi \langle \Xi^{n} \Sigma^{m+1} \rangle -
        {m\over V} \langle \Xi^{n} \Sigma^m \rangle \,.
\label{rel-XiSigma}
\end{equation}
This relation immediately implies that
\begin{equation}
\langle \Xi^{n} \rangle = (\kappa N_\chi)^n \langle \Sigma^n\rangle
  - {1\over V} \sum_{m=1}^{n-1} m (\kappa N_\chi)^m
       \langle \Xi^{n-1-m} \Sigma^{m-1} \rangle\,.
\end{equation}
For even values of $n$, 
repeated use of relation (\ref{rel-XiSigma}) gives 
\begin{equation}
\langle \Xi^{n} \rangle = (\kappa N_\chi)^n \langle \Sigma^n\rangle
   + \sum_{m=0}^{n/2-1} {a_{nm}
   (\kappa N_\chi)^{m +n/2} \over V^{n/2-m}} \langle \Sigma^{2m}\rangle\,,
\label{general-rel_AppA}
\end{equation}
where $a_{nm}$ are numerical coefficients.
Explicitly we obtain
\begin{eqnarray}
\langle \Xi^{2} \rangle &=& (\kappa N_\chi)^2 \langle \Sigma^2\rangle
   - {\kappa N_\chi\over V}\,, \nonumber \\
\langle \Xi^{4} \rangle &=& (\kappa N_\chi)^4 \langle \Sigma^4\rangle
   - 6 {(\kappa N_\chi)^3\over V}  \langle \Sigma^2\rangle +
     3 {(\kappa N_\chi)^2\over V^2}\,, \nonumber  \\
\langle \Xi^{6} \rangle &=& (\kappa N_\chi)^6 \langle \Sigma^6\rangle
   - 15 {(\kappa N_\chi)^5\over V}  \langle \Sigma^4\rangle 
\nonumber \\ 
    &+& 
    45 {(\kappa N_\chi)^4\over V^2}  \langle \Sigma^2\rangle -
    15 {(\kappa N_\chi)^3\over V^3}  \,.
\end{eqnarray}
Relation (\ref{general-rel_AppA}) proves Eq.~(\ref{EQXiN}) in the
infinite-volume limit. Size corrections decay as $1/V$.

It is easy to generalize this expressions to correlation functions.
For each point of the dual lattice $\widetilde{\bm x}$ we define the
local condensate
\begin{equation}
\Xi_{\widetilde{\bm x}} = {1\over 8} \sum_c 
\sum_{\langle {\bm x} \widetilde{\bm x}\rangle} 
\bar\chi^c_{\bm x} {\chi}^c_{\bm x}\,,
\end{equation}
where the sum is over the eight lattice points ${\bm x}$ that surround
the dual-lattice point. Relation (\ref{AppA-relz}) becomes
\begin{equation}
\langle \Xi_{\widetilde{\bm x}} f\rangle = 
  \kappa N_\chi \langle \sigma_{\widetilde{\bm x}} f\rangle -
   \langle \delta_{\widetilde{\bm x}} f \rangle\,.
\end{equation}
If we now take 
\begin{equation}
   f = \Xi_{\widetilde{\bm x}_1} \ldots
       \Xi_{\widetilde{\bm x}_n} 
       \sigma_{\widetilde{\bm x}_{n+1}} \ldots
       \sigma_{\widetilde{\bm x}_{n+m}} \,,
\end{equation}
and proceed as before, we obtain the local analogue of
Eq.~(\ref{general-rel_AppA}). If all points are distinct, i.e., we
disregard contact contributions, we have simply
\begin{equation}
\langle  \Xi_{\widetilde{\bm x}_1} \ldots
         \Xi_{\widetilde{\bm x}_n} \rangle = 
  \kappa^n N_\chi^n \langle  \sigma_{\widetilde{\bm x}_1} \ldots
         \sigma_{\widetilde{\bm x}_n} \rangle\,.
\end{equation}

\section{Monte Carlo simulations}
\label{app_montecarlo}

We simulate the lattice model with Hamiltonian~\eqref{KSmod} using the
hybrid MC algorithm~\cite{HKK-92, GLTRS-87-QCD}.  The
fundamental fields are $N_\chi=N_f/2$ (real) bosonic fields $\phi_{\bm
  x}^c$ defined on the lattice sites, the scalar field
$\sigma_{\widetilde{\bm x}}$ and its conjugate momentum
$\Pi_{\widetilde{\bm x}}$, defined instead on the dual
lattice~\cite{DKPR-87}. The hybrid MC Hamiltonian is
\begin{equation}
\begin{aligned}
  H_{\rm HMC} = &\sum_{{\bm x},{\bm y}}\sum_{c=1}^{N_\chi}\frac{1}{2}
  \phi_{\bm x}^c\big(M M^t\big)^{-1}_{{\bm x}, {\bm y}}\phi_{\bm y}^c\\
  &+ \frac{\kappa N_\chi}{2}\sum_{\widetilde{\bm x}}
  \sigma^2_{\widetilde{\bm x}}+\frac{1}{2}\sum_{\widetilde{\bm x}}
  \Pi^2_{\widetilde{\bm x}}\,.
\end{aligned}
    \label{def_Hhmc}
\end{equation}
For even values of $N_\chi$, this formulation is equivalent to the
original one with Hamiltonian~\eqref{KSmod}.  Indeed, the integration
of the fields $\phi_{\bm x}^c$ provides a factor $[\det
  (MM^t)]^{N_\chi/2} = \abs{\det M}^{N_\chi}$, and therefore
Eq.~\eqref{partition_only_sigma}.  Note that for odd $N_\chi$ this
algorithm does not sample the correct probability distribution of the
staggered fermions lattice system because of the presence of a sign
problem~\cite{HKK-92}.

In the simulations, we use a second-order minimum-norm symplectic
integrator for the update of the scalar field $\sigma_{\bm x}$ (the
integrator 2MN, as defined in Ref.~\cite{TF-05}).  We divide each
hybrid MC trajectory into four elementary integration steps, whose
length has been chosen in such a way that the acceptance is
approximately equal to 0.8.  Note that this prescription fixes the
number of inversions required to evaluate a single trajectory to $4
N_\chi$. The total lengths of the hybrid MC trajectories obtained have
an overall length of approximately $1.2-1.6$, depending on the lattice
size considered (the larger the size, the smaller the integration step
and the trajectory).  We observe that the average number of conjugate
gradient iterations required for a single inversion increases
approximately as $L$ for fixed inversion accuracy. As also reported in
the paper, the algorithm is subject to a severe slowdown for large
volumes.  The computer time required to obtain results with the same
uncertainty increases approximately as $L^6$ at the critical point
(see Ref.~\cite{Kennedy-99} for a general discussion of the efficiency
of the hybrid MC method).

We perform a measurement of the observables after each hybrid MC
update. Indeed, since most of the computer time is spent in the
update, especially for large values of $N_\chi$, the increase of the
frequency of the measurements does not have any significant impact on
the simulation time. To compute errors, we used standard blocking and
jackknife techniques.  Binnings of $10^3$ measures were always
sufficient to decorrelate completely our data.  The statistics
collected for the largest sizes are of the order of $1.8 \times 10^6,
8\times 10^5, 2.4\times10^6, 5.6\times 10^5$ measures, for $(L=30,
N_f=4)$, $(L=40, N_f=8)$, $(L=30, N_f=12)$, and $(L=40, N_f=16)$,
respectively.

\section{Parametrization of some universal FSS curves}
\label{interpolation}

In this appendix, we report the interpolation of the universal FSS
curves of the Binder parameter $U_4$ versus $R_\xi=\xi/L$,
cf. Eq.~(\ref{rscal2}), i.e. $U_4=F_U(R_\xi)$, for the available value
of $N_f$. In all cases the precision is approximately 0.5\% in the
considered interval.

For $N_f=16$, the interpolation of the numerical data for the largest
lattice sizes (for $L\ge 16$ there is no evidence of scaling
corrections) is given by
\begin{eqnarray}
  F_U(x) &\approx &
  2.99875 + 0.37513\,x +
  1.72310\,x^2 - 28.55675\,x^3
  \nonumber\\
  && + 62.07711\,x^4 - 61.03640\,x^5 +
         30.63879\,x^6         \nonumber\\
  && - 7.23447\,x^7 +
         0.54368\,x^8\,,
\label{fun16}
\end{eqnarray}
which reproduces the large-$L$ behavior of the data in the range
$0.35\lesssim x \lesssim 1.3$.

For $N_f=12$, an analogous procedure yields
\begin{eqnarray}
F_U(x) &\approx&
    3.04445    -   1.84745 x + 
   21.69942x^2 - 111.26622 x^3 
\nonumber \\
&& + 241.48614x^4 - 280.85629 x^5 + 
  184.32165x^6 \nonumber \\
&& -  64.48946 x^7 + 9.36899x^8,
\end{eqnarray}
which is again valid in the interval $0.35\lesssim x \lesssim 1.3$.
  
For $N_f=8$, we obtain (expression valid for $0.3 \lesssim x \lesssim
1.3$)
\begin{eqnarray}
  F_U(x) &\approx&
2.97785 + 0.67404\,x +
         0.57176\,x^2 - 39.62921\,x^3 
  \nonumber\\
  && + 108.87310\,x^4 - 135.91181\,x^5 + 91.11470\,x^6 
  \nonumber\\
  && - 31.88431\,x^7 +
         4.58106\,x^8\,,
\label{fun8}
\end{eqnarray}
while, for $N_f=4$, we have (for  $0.35 \lesssim x \lesssim
0.9$)
\begin{eqnarray}
  F_U(x) &\approx&
2.99015  + 1.09675\,x -
         6.49461\,x^2 - 6.12838\,x^3 
  \nonumber\\
  && - 16.68904\,x^4 + 173.36003\,x^5 -
         334.82353\,x^6
  \nonumber\\
  && + 266.46361\,x^7 - 78.55113\,x^8\, .
\label{fun4}
\end{eqnarray}

\end{document}